\documentclass[aps,revtex,prm,twocolumn,amsmath,superscriptaddress,showpacs,floatfix,reprint]{revtex4-2}
\usepackage{amsmath, nccmath}
\usepackage{amssymb}
\usepackage{bm}
\usepackage{braket}
\usepackage{natbib}
\usepackage{leftidx}
\usepackage{graphicx}    
\usepackage{verbatim}   
\usepackage{color}      
\usepackage{subfigure}  
\usepackage{hyperref}   
\usepackage{dcolumn}    
\usepackage{textcomp}
\usepackage{float}
\usepackage{threeparttable}
\usepackage{titlecaps}
\usepackage{multirow}
\usepackage{lipsum}
\hypersetup{
	colorlinks=true,
	citecolor=blue,
	filecolor=black,
	linkcolor=blue,
	urlcolor=cyan
}
\usepackage{mathastext}
\usepackage{mathptmx}
\usepackage{enumitem}
\DeclareGraphicsExtensions{.png,.pdf,.tif}
\usepackage{pict2e}
\usepackage[dvipsnames]{xcolor}
\usepackage[normalem]{ulem}

\begin{document}

\title{Effect on the Electronic and Magnetic Properties of Antiferromagnetic Topological Insulator MnBi$_2$Te$_4$ with Sn Doping}

\author{Susmita Changdar}
\affiliation{Condensed Matter Physics and Material Sciences Department, S. N. Bose National Centre for Basic Sciences, Kolkata, West Bengal-700106, India.}

\author{Susanta Ghosh}
\affiliation{Condensed Matter Physics and Material Sciences Department, S. N. Bose National Centre for Basic Sciences, Kolkata, West Bengal-700106, India.}
%
\author{Kritika Vijay}
\affiliation{Synchrotrons Utilization Section, Raja Ramanna Centre for Advanced Technology, Indore 452013, India.}
\affiliation{Homi Bhabha National Institute, Training School Complex, Anushakti Nagar, Mumbai, 400094, India.}
\author{Indrani Kar}
\affiliation{Condensed Matter Physics and Material Sciences Department, S. N. Bose National Centre for Basic Sciences, Kolkata, West Bengal-700106, India.}
\author{Sayan Routh}
\affiliation{Condensed Matter Physics and Material Sciences Department, S. N. Bose National Centre for Basic Sciences, Kolkata, West Bengal-700106, India.}
\author{P. K. Maheswari}
\affiliation{Condensed Matter Physics and Material Sciences Department, S. N. Bose National Centre for Basic Sciences, Kolkata, West Bengal-700106, India.}
\author{Soumya Ghorai}
\affiliation{Condensed Matter Physics and Material Sciences Department, S. N. Bose National Centre for Basic Sciences, Kolkata, West Bengal-700106, India.}
\author{Soma Banik}
\affiliation{Synchrotrons Utilization Section, Raja Ramanna Centre for Advanced Technology, Indore 452013, India.}
\affiliation{Homi Bhabha National Institute, Training School Complex, Anushakti Nagar, Mumbai, 400094, India.}
\author{S. Thirupathaiah}
\email{setti@bose.res.in}
\affiliation{Condensed Matter Physics and Material Sciences Department, S. N. Bose National Centre for Basic Sciences, Kolkata, West Bengal-700106, India.}

\date{\today}

\begin{abstract}
 We thoroughly investigate the effect of nonmagnetic Sn doping on the electronic and magnetic properties of antiferromagnetic topological insulator MnBi$_2$Te$_4$. We observe that Sn doping reduces the out-of-plane antiferromagnetic (AFM) interactions in MnBi$_2$Te$_4$ up to 68\% of Sn concentration, and above the system is found to be a paramagnetic. In this way, the anomalous Hall effect observed at a very high field of 7.8 T in MnBi$_2$Te$_4$ is reduced to 2 T with 68\% of Sn doping.  Electrical transport measurements suggest that all compositions are metallic in nature, while the low temperature resistivity is sensitive to the AFM ordering and to the doping induced disorder.   Hall effect study demonstrates that Sn actually dopes electrons into the system, thus, enhancing the electron carrier density almost by two orders at 68\% of Sn. In contrast, SnBi$_2$Te$_4$ is found to be a p-type system. Angle-resolved photoemission spectroscopy (ARPES) studies show that the topological properties are intact at least up to 55\% of Sn as the Dirac surface states are present in the valance band, but in SnBi$_2$Te$_4$ we are unable to detect the topological states due to heavy hole doping. Overall, Sn doping significantly affects the electronic and magnetic properties of MnBi$_2$Te$_4$.

\end{abstract}

\maketitle
\section{Introduction}\label{1}

Introduction of magnetism into a topological insulator has been known to give rise to various exotic quantum phenomena such as Axion~\cite{PhysRevLett.58.1799,PhysRevB.81.245209,li2010dynamical} and Chern insulating states, \cite{doi:10.1126/science.1187485,doi:10.1126/science.1234414,PhysRevLett.110.196801}, Majorana zero modes \cite{RevModPhys.83.1057}, and Weyl semimetallic phase~\cite{PhysRevB.83.205101,PhysRevLett.112.096804,kuroda2017evidence}. However experimental realization of such peculiar systems has been quite challenging due to the lack of intrinsic magnetism in topological insulators. Earlier, magnetic ordering in the topological insulators, (Bi/Sb)$_2$Te$_3$,  was achieved by doping with the magnetic ions like Cr/V~\cite{doi:10.1126/science.1234414,chang2015high,doi:10.1126/science.1187485,okada2016terahertz} in order to induce quantum anomalous Hall insulating (QAHI) phase. In this way, a ferromagnetic (FM) order is induced to break the time reversal symmetry (TRS) in the system. When TRS breaks, the degeneracy of surface Dirac point is lifted and the gap opening is followed by the chiral edge states, responsible for the QAH effect as observed in these systems. However, so far this method showed QAHI states at very low temperatures ($<$ 1 K)~\cite{doi:10.1126/science.1234414,tokura2019magnetic}.

Interestingly, MnBi$_2$Te$_4$ was recently discovered to be one of such intrinsic topological insulator with spontaneous AFM order~\cite{otrokov2019prediction,Gong_2019,zeugner2019chemical,doi:10.1126/science.aax8156,PhysRevResearch.1.012011,PhysRevLett.122.206401,PhysRevMaterials.3.064202,PhysRevX.9.041038,PhysRevX.9.041040,doi:10.1126/sciadv.aax9989,PhysRevB.100.121103,YAN2022164327}, in which the layers of ferromagnetically ordered Mn ions are coupled antiferromagnetically along the $c$-axis. Here, the long range magnetic ordering breaks TRS but the combination of TRS ($\Theta$) and half-lattice translational operator ($T_{1/2}$) connecting the two nearest Mn layers, i.e. $S=\Theta T_{1/2}$ remains protected~\cite{PhysRevLett.122.206401}. As a result, MnBi$_2$Te$_4$ becomes a potential candidate for hosting intriguing topological properties. It was also found that MnBi$_2$Te$_4$ thin flakes exhibit Axion insulating state for even number of layers~\cite{liu2020robust}. Chern insulating state is noticed for odd number of layers at zero magnetic field~\cite{doi:10.1126/science.aax8156}. But at much higher magnetic fields even number of layers also show Chern insulating phase~\cite{liu2020robust}. Similar to even number layered thin flakes, bulk MnBi$_2$Te$_4$ shows magnetization, thus requiring a very high magnetic fields (7.8 T) to obtain the QAH effect~\cite{C9CP05634C}. One approach to see QAH effect in MnBi$_2$Te$_4$ at lower fields is by weakening the interlayer AFM exchange coupling by doping with nonmagnetic ion like Sn at the Mn site.

\begin{figure*}[!ht]
	\includegraphics[width=\linewidth, clip=true]{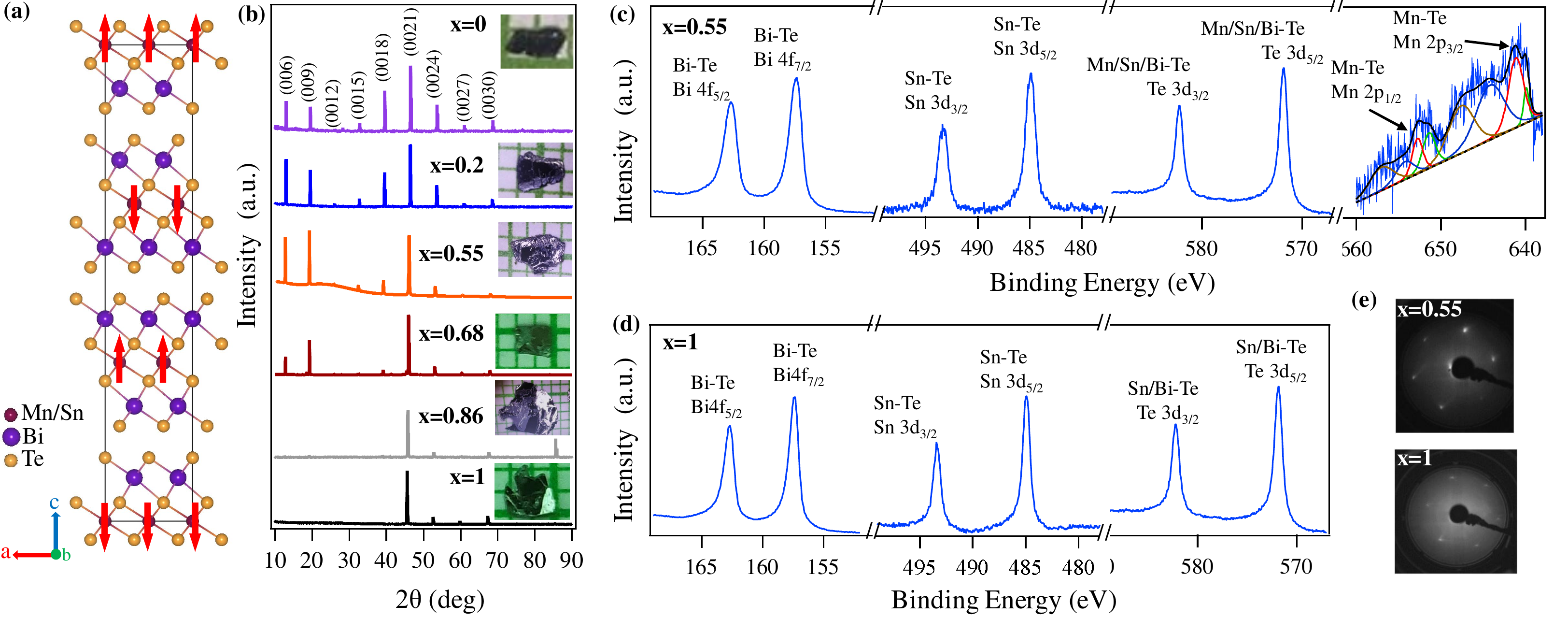}
	\caption{(a) Rhombohedral crystal structure of MnBi$_{2}$Te$_{4}$. Red arrows indicate the Mn$^{2+}$ ion spins in AFM state. (b) Single crystal XRD patterns showing $(00l)$ reflections from all the compositions.  Insets in (b) show single-crystal photographic images of each doping concentration on which the measurements were done. (c) and (d) Show XPS spectra exhibiting the Bi 4f, Sn 3d, Te 3d, and Mn 2p core levels for x=0.55 and Bi 4f, Te 3d, and Sn 3d core levels for x=1, respectively. (e) Shows LEED data taken with 70 eV photon energy revealing hexagonal pattern for both x=0.55 and x=1 single crystals.}
	\label{fig1}
\end{figure*}

In this work, we report comprehensively on the effect of Sn doping on the electronic and magnetic properties of the magnetic topological insulator MnBi$_2$Te$_4$. For this, we studied various single crystals of Mn$_{1-x}$Sn$_x$Bi$_2$Te$_4$ with x=0, 0.2, 0.55, 0.68, 0.86, and 1. We realized that Sn doping has a significant effect on the antiferromagnetic (AFM) ordering. That is, we observe a decrease in N\'eel temperature (T$_N$) from 24 K for MnBi$_2$Te$_4$ to 6 K for 68\% of Sn doping, while x=0.86 and 1 are found to be paramagnetic in nature. Further, we observe that the out-of-plane magnetic ordering is intact up to x=0.68, while for in-plane there is a change from linear M(H) behaviour to a very weak ferromagnetic type M(H) curve with a saturated magnetization of 0.8 $\mu_{B}$/f.u  with increasing Sn. We find a cusp on the resistivity data at the N\'eel temperature, which decreases with increasing Sn up to x=0.2, consistent with magnetic studies. However, as the Sn doping increased beyond x=0.2, we do not observe cusp at respective N\'eel temperatures but rather we observe low temperature resistivity upturn due to Sn doping induced disorder. On the other hand, SnBi$_2$Te$_4$ show Fermi-liquid type resistivity as the disordered is significantly reduced by fully replacing Mn with Sn. We further notice that Sn induces electron carriers into MnBi$_2$Te$_4$. As a result, the electron carrier concentration increase by two orders at a Sn concentration of 68\% compared to MnBi$_2$Te$_4$. In contrast to MnBi$_2$Te$_4$, SnBi$_2$Te$_4$ is found to be a p-type system. The Fermi level shift in the electronic band dispersions observed from ARPES studies on x=0.55 and x=1 single crystals is consistent with carrier concentrations estimated from the Hall measurements. We further notice that the topological properties are intact up to 55\% of Sn doping as the surface Dirac states are present in the valance band, while in SnBi$_2$Te$_4$ we are unable to detect them due to heavy hole doping.

\section{Experimental details}\label{2}

\begin{figure}[!ht]
	\includegraphics[width=1\linewidth, clip=true]{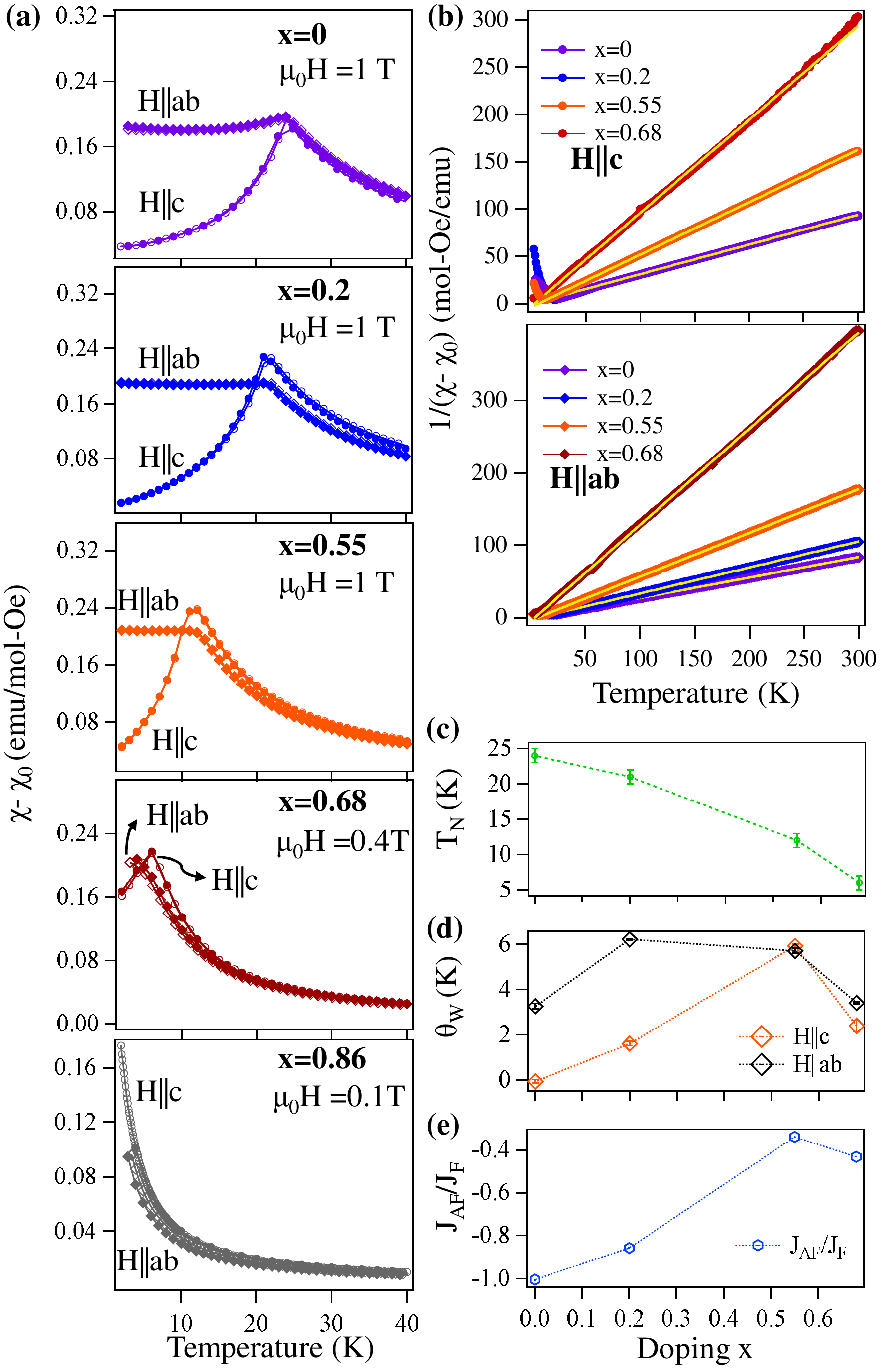}
	\caption{(a) Temperature dependent magnetic susceptibility ($\chi$) measured in field-cooled and zero field-cooled modes with H$\parallel$c and H$\parallel$ab. (b) Inverse magnetic susceptibility plotted as a function of temperature with  H$\parallel$c and H$\parallel$ab.  Doping dependent (c) N\'eel temperature (T$_{N}$), (d) Curie temperature ($\theta_{W}$) for H$\parallel$c and H$\parallel$ab, and (e) J$_{AF}$/J$_F$ for H$\parallel$c.}
	\label{fig2}
\end{figure}

We synthesized high quality single crystals of Mn$_{1-x}$Sn$_{x}$Bi$_{2}$Te$_{4}$ with x=0, 0.2, 0.55, 0.68, 0.86, and 1 using the solid state reaction route~\cite{C9CP05634C}. Constituent elements of Mn (99.9\%), Sn (99.99\%), Bi (99.99\%), and Te (99.99\%) were taken in stoichiometric ratio,  mixed thoroughly in argon atmosphere,  and kept in an alumina crucible before sealing it in a quartz ampoule under a vacuum of $10^{-4}$ mbar. The sealed quartz ampoule was then heated in a chamber furnace up to $900^{o}$C at a rate of $150^{o}$C/hr, and kept at that temperature for 24 hours. The ampoule was then cooled at a rate of $2^{o}$C/hr. The growth temperature of MnBi\textsubscript{2}Te\textsubscript{4} is $590^{o}$C and for SnBi$_2$Te$_4$ it is $600^{o}$C \cite{Orujlu_2020}. After the controlled cooling, we performed prolonged annealing within the temperature range of $590^{o}$C to $600^{o}$C for a maximum of seven days depending on the amount of Sn doping concentration. Finally, the ampoule was quenched in cold water. In this way, we got shiny and plate-like single crystals with a maximum size of 3$\times$3 mm$^2$.

As grown single crystals were structurally analyzed using the high-resolution X-ray diffractometer (Rigaku smartLab 9 kW) having Cu-K$_\alpha$ radiation. The chemical composition of the crystals were thoroughly examined using energy dispersive X-ray spectroscopy (EDS) on cleaved sample surfaces [see Fig.~S1 in supplemental information]. For convenience, herein, we will use respective nominal composition to represent each sample. Transport and magnetic properties were measured using Quantum Design Physical Property Measurement System (PPMS) with magnetic field up to 9 Tesla by varying the temperature between 2 and 300 K during the measurements.  XPS and ARPES measurements were performed at the undulator based Angle Resolved Photoelectron Spectroscopy beamline (ARPES BL-10), Indus-2, India. All samples were cleaved in-situ in the vacuum better than 5×10$^{–11}$ mbar and performed low energy electron diffraction (LEED) measurements using SPECS ErLEED 1000A to examine the sample surface quality. Both the XPS and ARPES measurements were carried out at a sample temperature of 20 K using SPECS Phoibos150 electron analyser. ARPES measurements were performed with a photon energy of h$\nu$ = 70 eV. The energy and angular resolutions were set at 40 meV and 0.1$^\circ$, respectively,  for the ARPES studies. XPS measurements were performed with a photon energy of h$\nu$= 800 eV and at an energy resolution of 0.4 eV.

\section{Results and discussion }\label{3}

MnBi$_2$Te$_4$ crystallizes in the rhombohedral structure with a space group of $R\bar{3}m$, consists  Te-Bi-Te-Mn-Te-Bi-Te septuple layers as shown in Fig.~\ref{fig1}(a). Fig.~\ref{fig1}(b) shows the XRD data of Mn$_{1-x}$Sn$_{x}$Bi$_2$Te$_4$ single crystals for x=0, 0.2, 0.55, 0.68, 0.86, and 1. XRD data of all crystals show clear reflections from $(00l)$ plane without any impurity peaks, confirming phase purity for all the crystals. Low energy electron diffraction (LEED) images taken on the freshly cleaved surfaces of x=0.55 and x=1 single crystals are shown in Fig.~\ref{fig1}(e). The LEED pattern of both samples are in hexagonal symmetry with clear first order diffraction spots coming from $(001)$ crystal planes. This is in good agreement with the XRD data, again confirming the high quality of single crystals. XPS data collected on x=0.55 and x=1 samples are shown in Figs.~\ref{fig1}(c) and ~\ref{fig1}(d), respectively. Both crystals exhibit major core level peaks at 157.35 eV of Bi 4f$_{5/2}$ and 162.75 eV of Bi 4f$_{7/2}$, which can be assigned to the Bi$^{3+}$ ions from the Bi-Te bond~\cite{doi:10.1063/5.0007440,C7RA08995C}. Additionally, we noticed two major peaks of Sn 3d$_{3/2}$ and Sn 3d$_{5/2}$ at 484.85 eV and 493.3 eV, respectively, from Sn$^{2+}$ ions associated to the Sn-Te bond~\cite{NEUDACHINA200577}. Te $3d$ core levels at 582.2 eV (Te 3d$_{3/2}$) and 571.85 eV (Te 3d$_{5/2}$) corresponding to the Te$^{2-}$ ions from Sn-Te, Mn-Te, and Bi-Te bonds of Mn$_{0.45}$Sn$_{0.55}$Bi$_2$Te$_4$~\cite{IWANOWSKI2004110} and also from Sn-Te, and Bi-Te bonds in SnBi$_2$Te$_4$. The peaks corresponding to Mn$^{2+}$ from Mn-Te bonds appear at 641.1 eV of Mn 2p$_{1/2}$ and 652.8 eV of Mn 2p$_{3/2}$~\cite{C9CP05634C,jiao2021layer}. We further noticed a few low intensity peaks related to Mn$^{2+}$  at 639.9 eV and 651.45 eV of the Mn-O bond as observed in a previous report on this system~\cite{C9CP05634C}.



\begin{figure*}[ht]
\includegraphics[width=0.99\linewidth, clip=true]{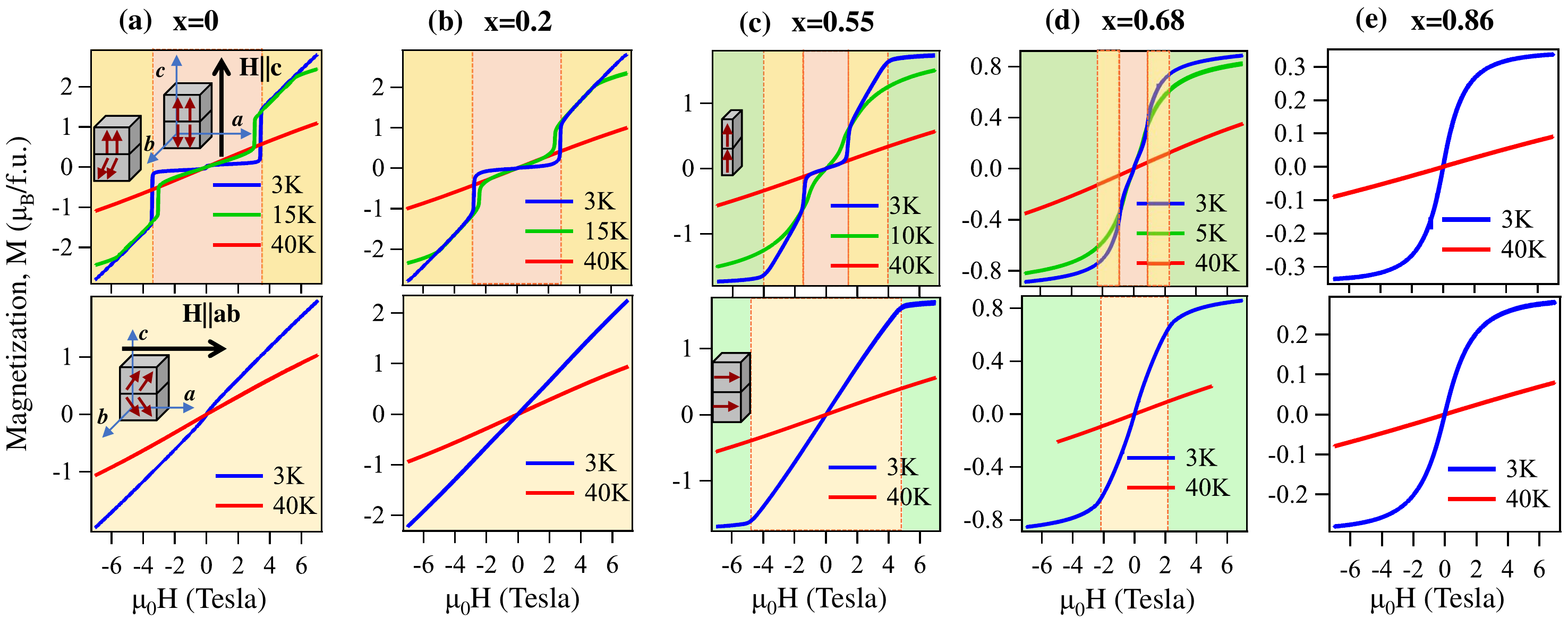}
\caption{Magnetization isotherms taken from (a) x=0, (b) x=0.2, (c) x=0.55,  (d) x=0.68, and (e) x=0.86 for the field applied parallel to $c$-axis (H$\parallel$c) (top figure) and parallel to $ab$-plane (H$\parallel$ab) (bottom figure) measured up to 7 T.  The insets are schematics representing the Mn spins in their respective color regions for two nearest Mn inter-layers (SLs).}
\label{fig3}
\end{figure*}

 To elucidate the effect of Sn doping on the magnetic properties of MnBi$_2$Te$_4$,  we performed thorough magnetic measurements on various Sn doping concentrations. Fig.~\ref{fig2}(a) depicts field-cooled (FC) and zero-field-cooled (ZFC) magnetic susceptibility ($\chi$) plotted as a function of temperature up to 40 K with magnetic field applied in-plane ($H\parallel ab$) and out-of-plane ($H\parallel c$) orientations. The FC and ZFC curves overlap for all samples, suggesting negligible magnetic hysteresis. In agreement with the previous reports~\cite{C9CP05634C, PhysRevResearch.1.012011}, MnBi$_2$Te$_4$ shows AFM ordering at a N\'eel temperature (T$_N$) of 24 K, thereby causing a sharp decrease in the magnetic susceptibility below $T_N$ for $H\parallel c$, whereas the magnetic susceptibility for $H\parallel ab$ remains almost constant below $T_N$. Next with increasing Sn substitution in Mn$_{1-x}$Sn$_x$Bi$_2$Te$_4$, the 2D ferromagnetic sublattice gets diluted, causing the decrease in $T_N$ linearly from 24 K for x=0 to 6 K for x=0.68 as shown in Fig.\ref{fig2}(c). The large decrease in $T_N$ implies a significant weakening of interlayer AFM coupling, while the overall AFM ordering is intact up to 68\% of Sn concentration. An earlier magnetic study suggested for a magnetic transition from AFM to paramagnetic for the 66\% of Sn doped Mn$_{1-x}$Sn$_x$Bi$_2$Te$_4$~\cite{PhysRevB.103.144407}. The discrepancy between our study and that of Ref.~\onlinecite{PhysRevB.103.144407} could be that the low temperature AFM ordering is dominated by the applied magnetic field when measured with 1 T [see Fig.~S2 in supplemental information]. Thus, we measured the magnetic data with a field of 0.4 T to see the AFM ordering at 6 K for x=0.68.

 Fig.~\ref{fig2}(b) shows the inverse magnetic susceptibility plotted as a function of temperature with $H\parallel c$ and $H\parallel ab$. From the Curie-Weiss law fitting to the data, we obtained Curie-Weiss temperature ($\theta_{W}$) for each composition and is plotted in Fig.~\ref{fig2}(d) as a function of Sn concentration. Earlier reports on MnBi$_2$Te$_4$ showed positive Curie-Weiss temperature for $H\parallel ab$  due to the in-plane FM exchange interaction and a negative Curie-Weiss temperature for $H\parallel c$ due to the out-of-plane AFM interactions~\cite{C9CP05634C}. In agreement to that, we obtained $\theta^{ab}_W$ = 3 K and $\theta^{c}_W$ = -0.08 K with an effective magnetic moment $\mu_{eff} = 5.3\mu_B/Mn$ for MnBi$_2$Te$_4$~\cite{zeugner2019chemical}. The estimated $\mu_{eff}$ value matches very well with that of Mn$^{2+}$ ions with spin $J=5/2$~\cite{C9CP05634C}. As shown in Fig.~\ref{fig2}(d),  we found different Curie-Weiss temperatures for in-plane ($\theta^{ab}_W$) and out-of-plane ($\theta^c_W$) field directions for the samples x=0 and 0.2, but at higher Sn concentrations both Curie-Weiss temperatures become almost equal due to decreasing magnetic anisotropy with increasing Sn concentration.

Moreover, both $\theta^{ab}_W$ and $\theta^c_W$ are found to be positive for x=0.2 and for other higher Sn concentrations, suggesting for  dominating FM interactions with increasing Sn. The strength of FM exchange interaction with respect to AFM can be determined from the ratio of $\theta_W/T_N$, which equals to $(J_{F}+J_{AF})/(J_{F}-J_{AF})$~\cite{coey_2010}. Here $J_F$ is intralayer FM exchange coupling and $J_{AF}$ represents interlayer AFM exchange coupling. The calculated $J_{AF}/J_{F}$ ratio is plotted in Fig.~\ref{fig2}(e) as a function of doping concentration. For MnBi$_2$Te$_4$, $J_{AF}/J_{F}$ is found to be -1, which indicates stronger AFM exchange coupling between two Mn-layers. Increase in Sn  concentration weakens the AFM coupling, thereby, decreases the $J_{AF}/J_F$ ratio as shown in Fig.~\ref{fig2}(e). Although the $J_{AF}/J_F$ ratio is relatively low for doped systems ($J_{AF}/J_F$=-0.43 for x=0.68) compared to the parent MnBi$_2$Te$_4$ ($J_{AF}/J_F$=-1 ), it is much higher than MnBi$_4$Te$_7$ [$J_{AF}/J_F \approx -0.04$]~\cite{PhysRevLett.124.197201}.

\begin{figure*}[ht]
\includegraphics[width=0.9\linewidth, clip=true]{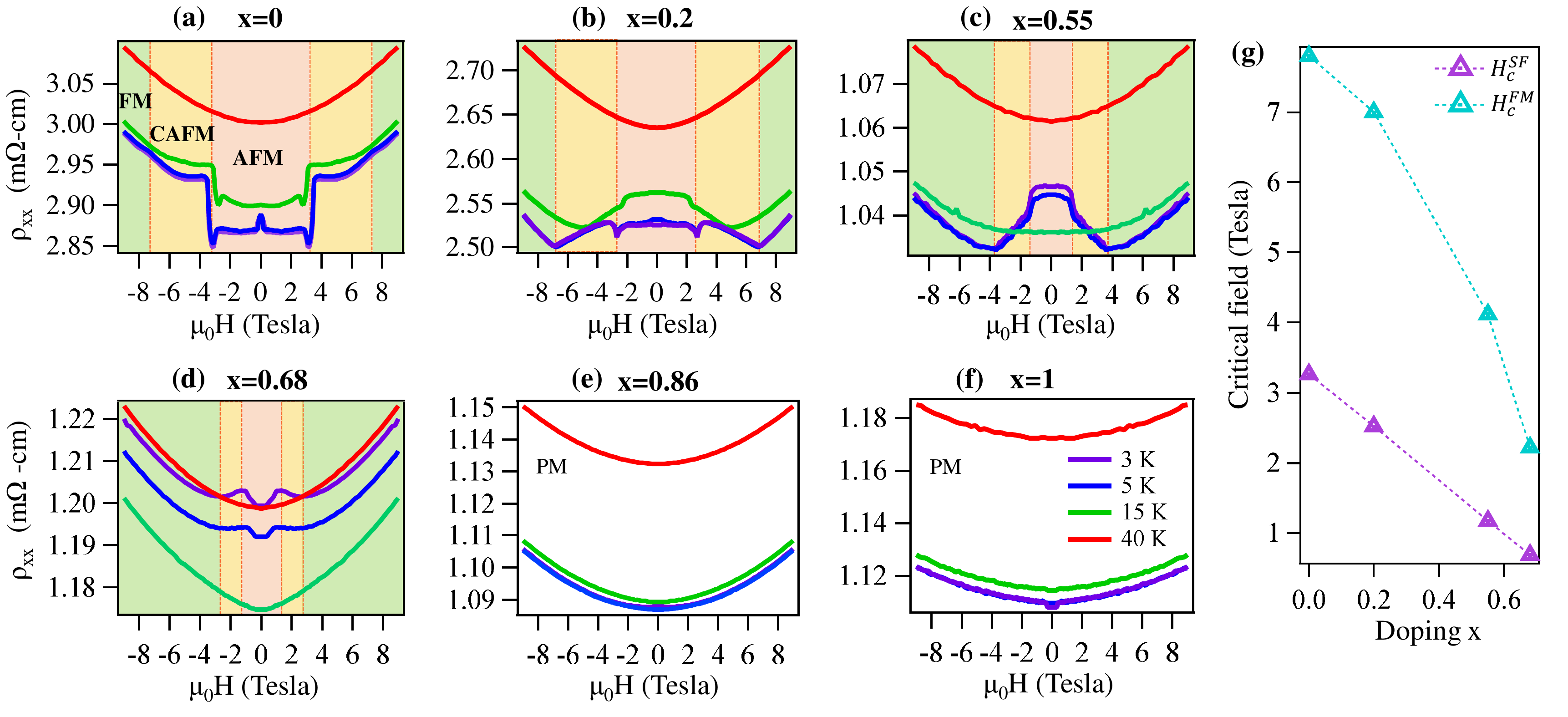}
\caption{ Field dependent magnetoresistance ($\rho_{xx}$) from (a) x=0, (b) x=0.2, (c) x=0.55, (d) x=0.68, (e) x=0.86, and (f) x=1 measured at 3 K, 5 K. 15 K, and 40 K. (g) Shows doping dependent critical fields of spin-flop transition (H$^{SF}_{c}$) and ferromagnetic ordering magnetic moment saturation (H$^{FM}_{c}$).}
\label{fig4}
\end{figure*}

\begin{figure}[ht]
\includegraphics[width=1\linewidth, clip=true]{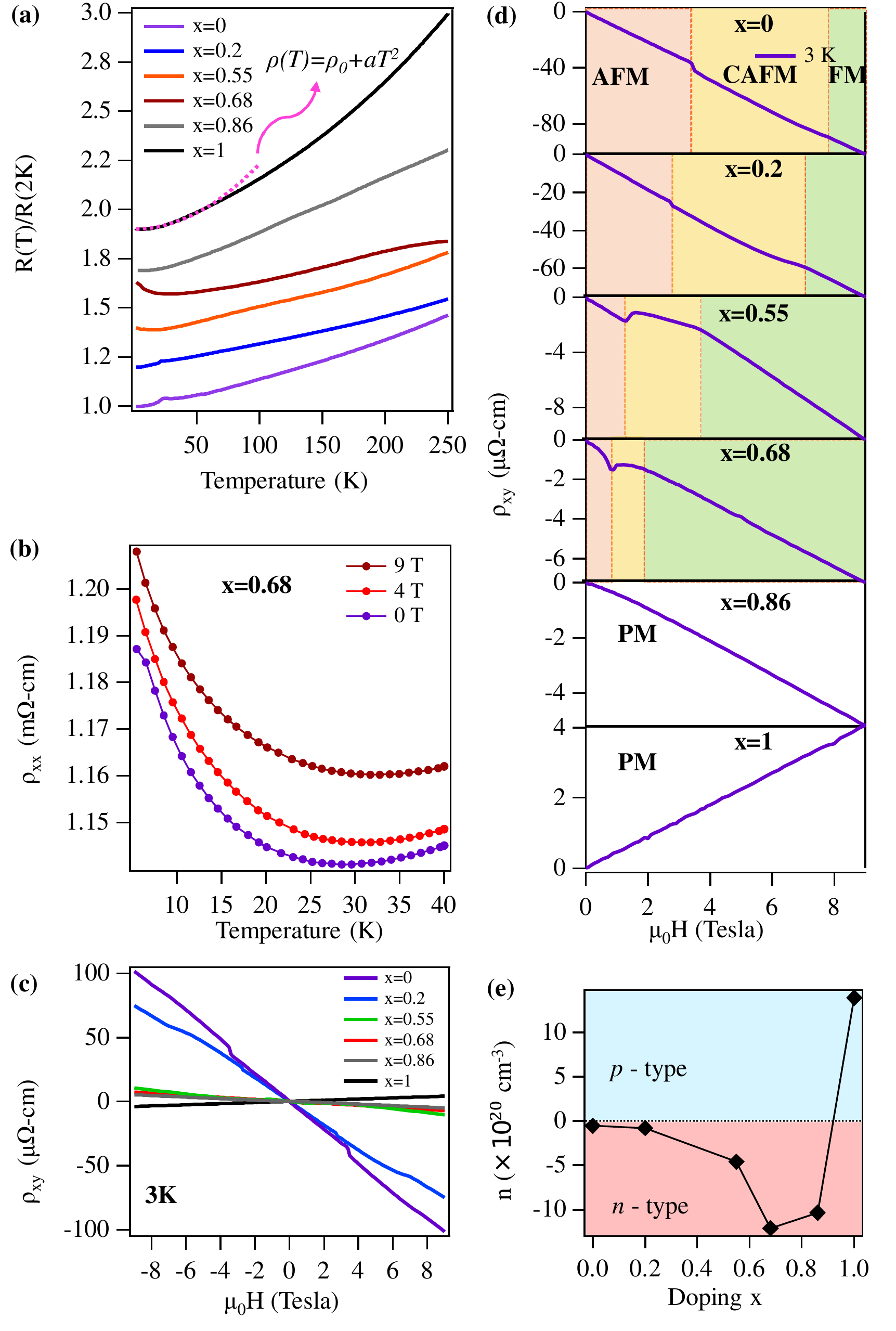}
	\caption{(a) Stacked in-plane resistivity ($\rho_{xx}$) plotted as a function of temperature from all the samples. Inset shows normalised in-plane resistivity (R(T)/R(2K)) vs. temperature curve at low temperature. (b) $\rho_{xx}$ vs. temperature data for x=0.68, measured with 0 T, 4 T, and 9 T applied magnetic field along crystallographic $c$-axis of the single crystal. (c) and (d) Field dependent Hall resistivity ($\rho_{xy}$) measured at 3 K. (red, yellow, green, and white colored regions in (d) signifies AFM, canted-AFM, FM, and PM states of the system, respectively.) (e) Charge carrier density (n) plotted as a function of Sn doping concentration.}
	\label{fig5}
\end{figure}

Figs.~\ref{fig3}(a)-(e) show field dependent magnetization M(H) isotherms measured with $H\parallel c$ and $H\parallel ab$. The crystallographic $c$-axis is known to be spin-easy axis for MnBi$_2$Te$_4$. In agreement to this, Fig.~\ref{fig3}(a) shows a spin-flop (SF) transition at a critical field of $H^{SF}_{c}$=3.3 T for $H\parallel c$ at 3 K, which gradually decreases to lower fields with increasing temperature up to the N\'eel temperature. Beyond the N\'eel temperature, the system transforms to paramagnetic (PM) in nature. During the spin-flop transition, the spin-structure of the system switches from an AFM state to a canted-AFM (CAFM) state, which upon further increase in magnetic field reaches to FM state due to complete polarization of the spins at a critical field of $H^{FM}_{c}$=7.8 T for MnBi$_2$Te$_4$~\cite{C9CP05634C}. With increasing Sn concentration, the spin-flop transition critical field decreases for $H\parallel c$. That means, for x=0.2 the critical field of spin-flop transition $H^{SF}_{c}$=2.5 T [see Fig.~\ref{fig3}(b)], which gradually goes down to $H^{SF}_{c}$=0.7 T for x=0.68 [see Fig.~\ref{fig3}(d)]. These observations further support our earlier argument that the AFM transition in x=0.68 sample can be detected only with the magnetic fields below 0.7 T. Further increase in Sn concentration to x=0.86, the spin-flop transition is completely disappeared and the system turns to be a very weak ferromagnet as it shows a sigmoid-like M(H) isotherm at 3 K for $H\parallel c$ [see Fig.~\ref{fig3}(e)].

Increase in Sn concentration not only decreases the spin-flop critical fields [$H^{SF}_{c}$], but also the saturation magnetization ($M_S$). MnBi$_2$Te$_4$ is known to show a saturation magnetization of 3.56 $\mu_B/f.u.$ at 8 T for $H\parallel c$~\cite{PhysRevB.100.104409}, it goes down to a negligible saturation magnetization of 0.32$\mu_B/f.u$ for x=0.86. In fact at 3 K, we notice magnetization saturation at 4 T for x=0.55 and at 2.2 T for x=0.68 due to weak FM ordering induced with Sn doping.  On the other hand, for $H\parallel ab$, the magnetization M(H) isotherms of x=0, 0.2, and 0.55  show linear dependents on H due to gradual polarization of spins towards the applied field direction. However, the magnetic saturation is noticed for x=0.55 at around 5 T. Interestingly, for x=0.68 and 0.86 the nature of the M(H) curves at lower fields resembles to a weak ferromagnet with sigmoid-like shape and magnetic saturation is around 2 T.  These observations suggest that Sn doping into MnBi$_2$Te$_4$ transforms the system to a weak ferromagnet.

Next, field dependent magnetoresistance (MR) up to 9 T measured at various temperatures for all compositions are plotted in Figs.~\ref{fig4}(a)-(f). During the measurements, the magnetic field was applied along the $c$-axis with current applied parallel to $ab$-plane. From Fig.~\ref{fig4}(a), we can see that $\rho_{xx}(H)$ of MnBi$_2$Te$_4$ jumps suddenly at a critical field of 3.2 T when measured below T$_N$, and again slope changes at a critical field of $7.8$ T. The critical fields found from MR data are matching very well with that of $H^{SF}_{c}$ and $H^{FM}_{c}$ from the M(H) data. We notice that the field dependent MR behaviour is consistent with M(H) data up to N\'eel temperature, and above T$_N$ we observe a parabolic field dependent MR, in agreement with an earlier  report~\cite{PhysRevLett.124.197201}.  Also, the critical fields decrease monotonically with increasing Sn as demonstrated in Fig.~\ref{fig4}(g).  Next, x=0.86 and x=1 (SnBi$_2$Te$_4$) show parabolic field dependent MR at all the measured sample temperatures due to their paramagnetic nature [see Figs.~\ref{fig4}(e) and ~\ref{fig4}(f)].

Temperature dependent in-plane resistivity ($\rho_{xx}$) data are plotted in Fig.~\ref{fig5}(a). $\rho_{xx}$ of all samples suggest an overall metallic behavior. We notice a cusp on the resistivity at 24 K for at x=0 and 21 K for x=0.2, due to the spin-fluctuations triggered at respective N\'eel temperatures.  On the other hand, the compositions with higher Sn such as x=0.55 and x=0.68 we find no effect of AFM ordering on $\rho_{xx}$ data at their respective $T_N$'s. But both these compositions show upturn resistivity with minima at 15 K for x=0.55 and at 28 K for x=0.68. Note here that these temperature minima are higher than their respective N\'eel temperatures, suggesting rather a different mechanism for increasing resistivity at low temperatures such as the disorder induced electron-electron ($e-e$) interactions~\cite{Chakraborty1996}, weak localization (WL)~\cite{Abrahams1979}, or the Kondo effect~\cite{Kondo1964}. In order to explore further on the origin of resistivity upturn, we performed field dependent resistivity on x=0.68 and plotted the data as shown in Fig.~\ref{fig5}(b). From Fig.~\ref{fig5}(b) we can see that up to 9 T the resistivity upturn does not suppress, contradicting to the possibility of Kondo effect or weak localization in which case the resistivity upturn suppresses~\cite{Kondo1964, Ciesielski2020}. Thus, low temperature resistivity upturn found could be originated from the induced disorder with Sn doping~\cite {doi:10.1021/acs.nanolett.9b01412, Jia2010, Chakraborty1996}.  Nevertheless, the resistivity upturn is absent from SnBi$_2$Te$_4$ compound and rather we find a clean Fermi-liquid type resistivity curve that is satisfying the T$^2$ law.

\begin{figure}[ht]
\includegraphics[width=0.95\linewidth, clip=true]{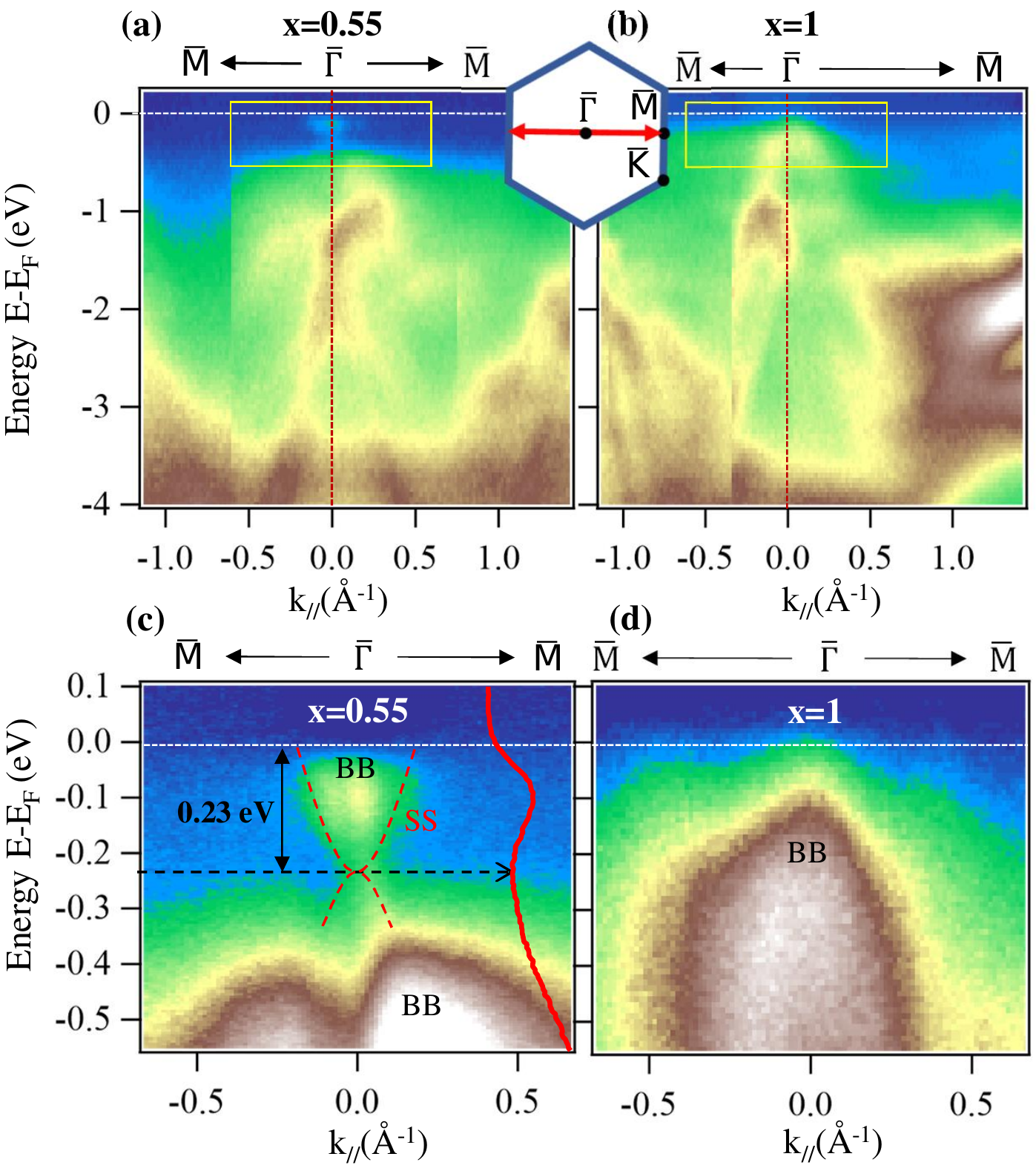}
\caption{(a) and (b) Energy distribution maps (EDMs) of x=0.55 and x=1, respectively,  taken along $\overline{\Gamma}-\overline{M}$ direction with 70 eV photon energy using linearly polarized light. (c) and (d) are the zoomed-in EDMs of x=0.55 and x=1 near $E_F$. The red curve in x=0.55 is the energy distribution curve taken by integrating around the $\overline{\Gamma}$ point.}
\label{fig6}
\end{figure}

Field dependent Hall resistivity ($\rho_{xy}$) measurements at 3 K for all compositions are shown in Fig.~\ref{fig5}(c). As can be seen from Fig.~\ref{fig5}(c), the $\rho_{xy}$(H) decreases with increasing Sn concentration up to x=0.68 and then slope reverses the sign for SnBi$_2$Te$_4$. The change in slope of $\rho_{xy}$(H) clearly hints at increase in electron carrier concentration up to x=0.68 and then the system SnBi$_2$Te$_4$ becomes a p-type when Mn is totally replaced by Sn. To explore the Hall resistivity in details, we plotted $\rho_{xy}$(H) individually for every composition as shown Fig.~\ref{fig5}(d). From a closer look on the Hall data of MnBi$_2$Te$_4$ we can notice a change in slope at the spin-flop critical field H$^{SF}_c$=3.2 T and the ferromagnetic ordering critical field H$^{FM}_{c}$= 7.8 T. The change in slope at the critical fields is due to the anomalous Hall effect. With increasing Sn concentration, the slope change in Hall resistivity appears at much lower fields of H$^{SF}_c$=0.8 T and H$^{FM}_c$= 2 T when measured on x=0.68 sample. These observations clearly demonstrate the tuning of anomalous Hall effect from higher fields to lower fields in going from parent to 68\% of Sn doping. However, we do not observe AHE from x=0.86 and x=1 samples as they are the paramagnetic in nature.

Next, the charge carrier concentration has been estimated from the Hall data and plotted in Fig.~\ref{fig6} (e) for all the compositions. We calculated an electron carrier density of $5.4\times10^{19}$ cm$^{-3}$ for x=0, $8.18\times10^{19}$ cm$^{-3}$ for x=0.2, $4.59\times10^{20}$ cm$^{-3}$ for x=0.55, $1.2\times10^{21}$ cm$^{-3}$ for x=0.68, and  $1.04\times10^{21}$ cm$^{-3}$ for x=0.86. Thus, in going from x=0 to x=0.68 the electron carrier density increased by an order of 2 and then starts decreasing. On the other hand, for x=1 (SnBi$_2$Te$_4$) we estimated a hole carrier density of $1.39\times10^{21}$cm$^{-3}$, which is in good agreement with previous reports on hole carrier concentration of SnBi$_2$Te$_4$ \cite{article1,zou2018atomic,TAK2017966}. Further, our observation of increase in electron density with Sn doping is in good agreement with a previous report~\citep{PhysRevB.103.144407}.

In order to understand the effect of Sn doping on the electronic band dispersions near the Fermi level,  we performed ARPES measurements on x=0.55 and x=1 single crystals as shown in  Fig.~\ref{fig6} measured with 70 eV photon energy of linearly polarized light. Figs.~\ref{fig6}(a) and ~\ref{fig6}(b) depict energy distribution maps (EDMs) plotted for x=0.55 and x=1, respectively,  along $\overline{\Gamma}-\overline{M}$ orientation as schematically shown in the inset. Figs.~\ref{fig6}(c) and ~\ref{fig6}(d) depict zoomed-in images of~\ref{fig6}(a) and (b), respectively, showing band dispersions near the Fermi level. From Fig.~\ref{fig6}(c), we can notice surface Dirac state with a band crossing point at a binding energy of 0.23 eV as usually observed in MnBi$_2$Te$_4$~\cite{Hao2019, PhysRevX.9.041040, Shikin2020}.   The band crossing point is estimated with the help of energy dispersive curve (EDC) extracted by integrating the EDM around $k_\parallel$=0, overlapped on the EDM. However, the surface Dirac state completely shifted above the Fermi level when Mn is totally replaced by Sn [see Fig.~\ref{fig6}(d)] due to Fermi level shifting towards higher binding energy. This is in good agreement with the dominant hole carrier density estimated from the Hall measurements in the case of SnBi$_2$Te$_4$. Overall, the band dispersions of MnBi$_2$Te$_4$ along $\overline{\Gamma}-\overline{M}$  orientation are in good agreement with previous reports~\cite{acsnano.1c03936}.  Though we are unable to resolve surface and bulk bands more clearly near the Fermi level within our experimental energy resolution (40 meV), the main message is clear that the topological Dirac surface states are intact at least up to 55\% of Sn doping in MnBi$_2$Te$_4$. On the other hand, the bulk band dispersions look almost identical between x=0.55 and x=1, except that the bulk bands shifted towards the Fermi level in x=1 compared to x=0.55.

Despite MnBi$_2$Te$_4$ being topological insulator with intrinsic magnetic ordering, there are mainly two obstacles on the path of realizing Chern insulating state in this system. One of them is the high magnetic fields needed to obtain FM order~\cite{Hao2019} and the other one is the dominant bulk electron carriers induced by the excess Bi and the antisite defects in single crystals~\cite{chen2019intrinsic}, shifting the Fermi level into the bulk conduction band. Doping Sn at the Mn site appeared to be a promising technique for suppressing the $n$-type bulk carriers as SnBi$_2$Te$_4$ is isostructural to MnBi$_2$Te$_4$ and has high $p-$type carrier concentration~\cite{zou2018atomic}. But the magnetism, an important ingredient to realize the Chern insulating state~\cite{Ranmohotti2013}, seems to be disappearing beyond 68\% of Sn doping while still the electron carriers dominating the transport. On the other hand, when the electron density decreased at higher Sn doping (x=0.86) the system turns into nonmagnetic [see Fig.~\ref{fig5}].

\section{Conclusions}\label{4}

In conclusion, we have successfully grown high quality single crystals of Mn$_{1-x}$Sn$_{x}$Bi$_2$Te$_4$ (x=0, 0.2, 0.55, 0.68, 0.86 and 1). We noticed that Sn doping in MnBi$_2$Te$_4$ is an effective way of acquiring quantum Hall state at lower magnetic fields as the critical field of FM ordering goes from 7.8 T to 2 T in going from x=0 to x=0.68. Electrical resistivity is found to be sensitive to the AFM ordering temperature (T$_N$), but with Sn doping the low temperature resistivity shows upturn due to the doping induced disorder. However, low temperature upturn is absent in SnBi$_2$Te$_4$ as the disorder is reduced. Hall effect study shows electron doping into the system with Sn, and thus, enhancing the electron carrier density almost by two orders with 68\% of Sn concentration. In contrast, SnBi$_2$Te$_4$ is found to be a $p$-type system. ARPES studies show that topological properties are intact up to 55\% of Sn doping, but  the Dirac surface states disappeared in SnBi$_2$Te$_4$ as they moved to conduction band due to heavy hole doping. Thus, our studies clearly demonstrate the effect of Sn doping on the electronic and magnetic properties of MnBi$_2$Te$_4$.

\section{Acknowledgement}\label{5}

S.C. acknowledges University Grants Commission (UGC), India for the Ph.D. fellowship. S.T. acknowledges financial support by the Department of Science and Technology (DST) through the grant no. SRG/2020/000393. Dr. Tapas Ganguli from RRCAT is thanked for all the support and encouragement.
\bibliography{MnBi2Te4}

\end{document}